# A Novel Framework Integrating AI Model and Enzymological Experiments Promotes Identification of SARS-CoV-2 3CL Protease Inhibitors and Activity-based Probe

Fan Hu[1,#], Lei Wang[2,#], Yishen Hu[1], Dongqi Wang[1], Weijie Wang[2], Jianbing Jiang[3], Nan Li[2,*] and Peng Yin[1,*]

[1]Guangdong-Hong Kong-Macao Joint Laboratory of Human-Machine Intelligence-Synergy Systems, Shenzhen Institutes of Advanced Technology, Chinese Academy of Sciences, Shenzhen, 518055, China, [2]CAS Key Laboratory for Quantitative Engineering Biology, Shenzhen Institute of Synthetic Biology, Shenzhen Institutes of Advanced Technology, Chinese Academy of Sciences, Shenzhen, 518055, China and [3]School of Pharmaceutical Sciences, Shenzhen University Health Science Center, Shenzhen, 518055, China.

#: These authors contributed equally to this work.

*To whom correspondence should be addressed: nan.li@siat.ac.cn, peng.yin@siat.ac.cn

# Abstract

The identification of protein-ligand interaction plays a key role in biochemical research and drug discovery. Although deep learning has recently shown great promise in discovering new drugs, there remains a gap between deep learning-based and experimental approaches. Here we propose a novel framework, named **AIMEE,** integrating **AI M**odel and **E**nzymology **E**xperiments, to identify inhibitors against 3CL protease of SARS-CoV-2, which has taken a significant toll on people across the globe. From a bioactive chemical library, we have conducted two rounds of experiments and identified six novel inhibitors with a hit rate of 29.41%, and four of them showed an $IC_{50}$ value less than 3 μM. Moreover, we explored the interpretability of the central model in AIMEE, mapping the deep learning extracted features to domain knowledge of chemical properties. Based on this knowledge, a commercially available compound was selected and proven to be an activity-based probe of $3CL^{pro}$. This work highlights the great potential of combining deep learning models and biochemical experiments for intelligent iteration and expanding the boundaries of drug discovery.

# Introduction

Drug discovery is one of the most powerful weapons in the fight against diseases. Due to recent changes in human behavior such as globalization and an increasing stress placed on the natural environment, the arms race between humanity and disease has intensified. The ongoing COVID-19 pandemic has so far sickened more than one hundred million and killed over two million people across the globe as of January, 2021[1], and it's been just over a year since the SARS-CoV-2 was reported[2–4]. Therefore, it is imperative to be constantly updating our technology to address challenges posed by existing and possible new emerging diseases. For SARS-CoV-2, the viral main protease ($M^{pro}$ or $3CL^{pro}$) is an attractive drug target for COVID-19 drug discovery, given its essential role in viral infection and there being no human homologue[5–9]. However, so far, no specific drugs against SARS-CoV-2 have been approved. Considering high-throughput screening is not available in most biological safety level 3 laboratories (where SARS-CoV-2 cell-based assays can be performed), a feasible and cost-effective way is to combine in vitro experiments with in silico screening.

Deep learning has recently been applied successfully in many fields including drug discovery[10–13]. The deep learning based method may combine the advantages of structure-based and ligand-based drug design methods and lead to superior performance[14–18]. More importantly, the rapidly increasing amount of data about SARS-CoV-2 enables deep learning to efficiently extract useful features and thus significantly improve the prediction accuracy. There have been several applications of deep learning models for screening inhibitors targeting important viral proteins such as $3CL^{pro}$ and the spike protein since the outbreak of COVID-19[19–22], although most of these models did not verify their predictions in vitro. Presently, it is hard to judge whether traditional methods or AI-aided methods are better suited to practical applications of drug discovery. Besides the quantity and quality requirements of the data, deep learning-based methods rely highly on appropriate design of experiments and might suffer from the generalizability issue. Thus the gap between deep learning based methods and experimental approaches remains. Ideally, one of the most promising ways is to combine deep learning models and biochemical experiments to build a closed and iterative loop, enabling the model to constantly learn and boosting its accuracy for the target task[23].

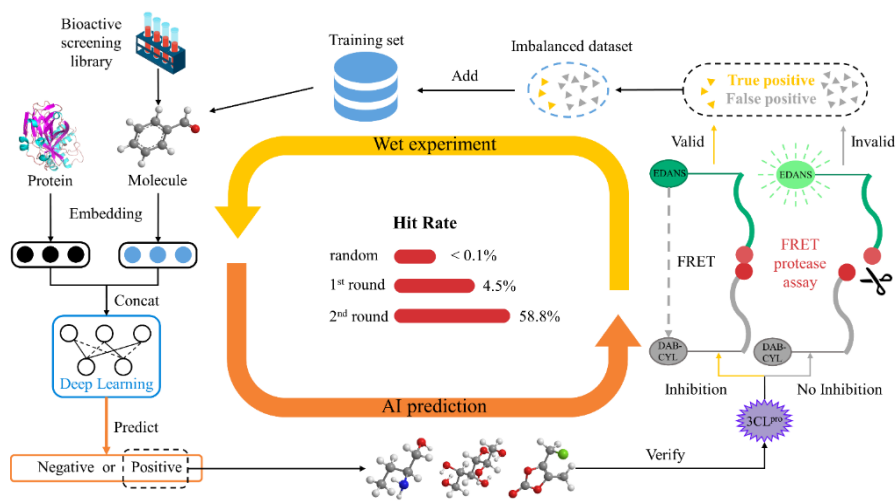

Fig. 1 Schematic of the proposed framework and the hit rates in the closed-loop. The framework consists of two parts: model training and enzymological experiments. After several rounds of iteration, the screening hit rate increased.

In the present study, we propose a novel framework integrating an **AI M**odel and **E**nzymological **E**xperiments (**AIMEE**) to identify inhibitors against the SARS-CoV-2 3CL protease. Our framework involves three stages. First, we pre-trained our model with the necessary information including protein sequences, protein structures and protein-ligand interactions. The model has achieved the top performance on a benchmark dataset over all existing scoring functions. Second, we leveraged the model on a carefully curated imbalanced 3CL$^{pro}$ inhibitor dataset and then applied the resulting model to screen a chemical library including approved, clinical-stage drugs and bioactive compounds. Next, compounds with high predicted probability were selected for verifying their binding affinities with 3CL$^{pro}$ *in vitro*. Third, we integrated the new experimental data and updated the model according to the experimental results. Finally, we iterated steps two and three.

Through this approach we identified several 3CL$^{pro}$ inhibitors, including six inhibitors with the IC$_{50}$ value less than 20μM. Importantly, some inhibitors showed potential clinical value treating COVID-19. For example, bacitracin, which shows the inhibition of 3CL$^{pro}$ with an IC$_{50}$ of 1.353 μM, was approved for intramuscular injection in the treatment of staphylococcal pneumonia and empyema in infants. Remarkably, the 100 μM inhibitor primary screening hit rate rose from 4.46% in the 1$^{st}$ round to 58.82% in the 2$^{nd}$ round and the strong binding (IC$_{50}$ value < 20 μM) hit rate rose from 0.13% to 29.41% (Fig. 1), which suggests a significantly increased accuracy of the model during the process. Computationally, a highly imbalanced dataset with numerous inactive compounds against a limited number of active compounds could lead to a negative impact on the model performance. We demonstrated that our method reduces this negative effect by leveraging various techniques, which could be introduced into many drug discovery applications. Furthermore, we demonstrated the binding positions of the identified inhibitors, explored the logic behind the model and evaluated how the model discerns the key sites of the identified compounds. Based on this biological interpretation, we showed a commercially available compound to be activity-based probe (ABP)[24], which could be used for the activity-based protein profiling (ABPP) of the target protein to study its enzymological properties and functions during infection[25]. This work highlights a promising prospect of uniting deep learning models and biochemical experiments for intelligent iteration and expanding the boundaries of drug discovery.

# Results and Discussion

## Model pre-training and performance

Among recently proposed deep learning-based methods, structure-based methods utilizing the strong feature extraction capability of deep learning to capture information from protein-ligand complexes, showed a huge improvement compared to docking methods. However, these methods are limited by the lack of data and cannot extract the same information from sequence-only data, which are far more abundant than structural data. On the other hand, sequence-based methods may be disadvantaged by incomplete sample information. Here we designed a graph enhanced transformer model to predict protein-ligand interactions, which combines the advantages of both structure-based and sequence-based methods.

We explored multi-modal inputs of proteins to capture information from different dimensions. At first, the modified Transformer[26] was pre-trained using large scale unlabeled protein sequences, which extracted protein evolutionary relationships using self-supervised learning. Then, for a protein with solved structural data, we utilized the pretrained Transformer and a graph attention network (GAT) to process its sequence and structure, respectively, to get the protein embedding vector. The corresponding drug molecule was processed by GAT to get the drug embedding vector. Next, these two representation vectors were concatenated and fed into a linear regression layer to predict their binding (Fig. 2a).

We evaluated the model on the diverse 290 complexes within PDBbind v.2016 core set and split the training and validation sets in the same manner as Pafnucy[16]. Pearson's correlation coefficient R and root mean square error (RMSE), which refer to the linear correlation and the differences between the predicted and real values, respectively, were used as evaluation metrics. Our model achieved RMSE=1.274 and R=0.818 on the PDBbind v.2016 core set. As shown in Fig. 2b, our model has achieved the top performance on this structural benchmark dataset over existing methods including docking and deep learning based methods.

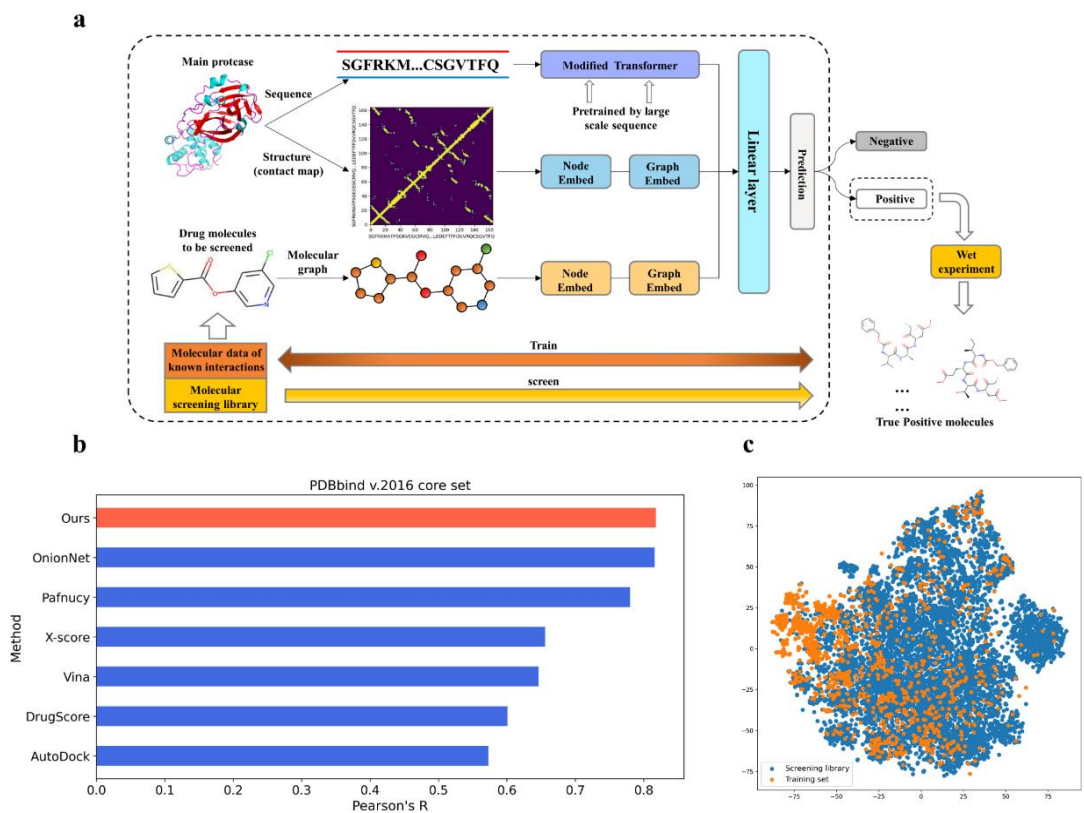

Fig. 2 Model training and performance evaluation. **a** The architecture of the proposed model. **b** Our model has achieved the top performance on the benchmark PDBbind v2016 set. Pearson's correlation coefficient (R), which is used to evaluate the linear correlation between the predicted and real values, is the most important measure for evaluating this benchmark dataset. **c** Dimensionality reduction by t-SNE of the molecule representations of screening library (blue) and training set (orange).

## Deep learning based screening

One of the greatest advantages for deep learning is its ability to constantly learn and improve its accuracy for a specific task. Here, to obtain novel 3CL$^{pro}$ inhibitors in an efficient and inexpensive way, we have built a closed-loop combining our deep learning model and wet lab experiments and have conducted two rounds of experiments.

In the first-round of experiments, a curated dataset (dataset v1) containing 304 3CL$^{pro}$ inhibitors with their binding affinities was used to train our model. Among these, 160 compounds with IC$_{50}$ less than 20 μM were regarded as positive samples whereas another 144 compounds were regarded as negative samples. After training, our model selected 740 candidate compounds from a bioactive chemical library of 10,924 compounds consisting of approved, clinical-stage and bioactive compounds. Among these candidates, we verified 33 compounds which showed an inhibitory effect at 100 μM in primary screening and then we calculated their IC$_{50}$ values. Using IC$_{50}$ of 20 μM as a hit cut-off, we identified one positive compound (IC$_{50}$ = 1.353 μM) and 739 negative compounds for 3CL$^{pro}$. Then we incorporated these results and data carefully curated from other sources to build the 3CL inhibitor refined set (dataset v2). We have also expanded the negative set, which contains almost 300,000 negative compounds for SARS-CoV/SARS-CoV-2 3CL$^{pro}$, for negative sample sampling, as described in the material and methods section.

In the second-round of experiments, the refined set consisting of 408 positive and 1,859 negative compounds was used to re-train our model. At first, we tried to re-train our model directly on the merged set of the refined set and the negative set (i.e., 408 positive and over 300,000 negative compounds) to fully extract features from lots of negative compounds. However, the performance metrics were very low, indicating a negative impact on the model training. This high data imbalance seems to be a very common problem in bioassay datasets. A recent study showed this extreme imbalance issue results in very poor performance, especially for machine learning algorithms (e.g., average precision, recall and mcc of 0.012, 0.53 and 0.056, respectively, for a set consisting of 1,181 positive and 256,343 negative compounds), although several strategies have been tried to mitigate this imbalance[27].

To ensure the reliability of the screening results, we explored a strategy to improve model performance. First, we randomly selected 80,000 samples from the negative set and merged them with the refined set. Then, the model was trained on the merged set using three-fold cross validation, which means each fold contains a similar number of positive and negative samples. To test model sensitivity for different negative samples, we repeated the experiment three times by randomly choosing 80,000 negative samples from the negative set. During training, a focal loss[28], which focuses more on the minority class and hard samples, was used instead of binary cross entropy loss. At last, our model achieved average precision, recall and mcc of 0.716, 0.515 and 0.605, respectively, on this highly imbalanced dataset (Supplementary Table 1). These evaluation metrics indicated that for each fold consisting of 136 positive and 27,286 negative samples, our model had predicted 98 positive samples in which 70 are true positive. This was a great improvement on such an imbalanced dataset and it is of important significance for drug discovery, especially for *in silico* screening, because biochemical experiments usually produce large numbers of negative but a small fraction of positive samples (e.g., 0.01% to 0.14% typical hit rate for a high throughput screening). We also tested different numbers of negative sampling and the results indicated that the 80,000 samples were appropriate considering various parameters including accuracy, recall, mcc and computational cost. Furthermore, the sensitivity analysis by three rounds of random down-sampling and three-fold cross validation verified the stability of the model. Lastly, we used these models to screen the bioactive chemical library after excluding the refined set. For each down-sampling group, the compounds with predicted probability higher than 0.5 in at least two folds were curated. A total of 17 compounds were selected as the final candidates.

## In vitro screening for 3CL[pro] inhibitors.

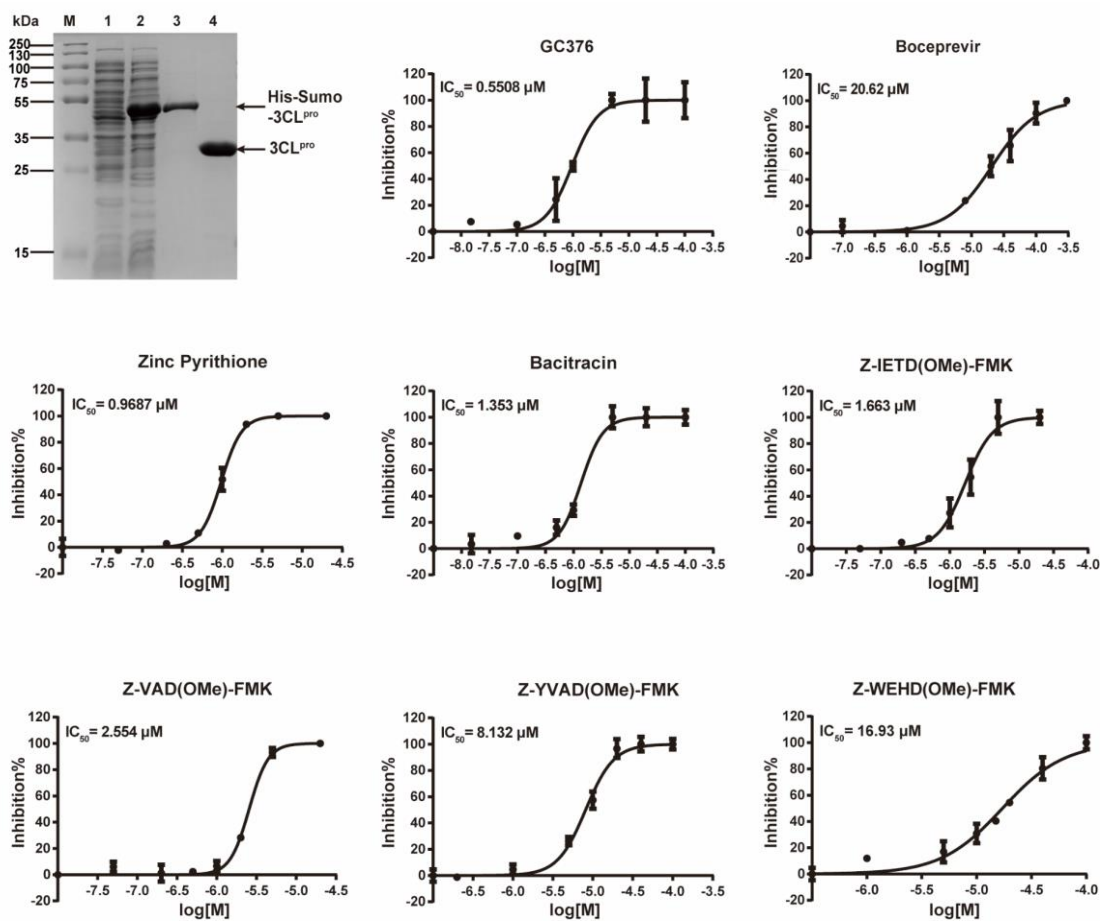

Fig. 3 Purification of 3CL[pro] and enzyme inhibitory activity of the inhibitors. **a** Purification of 3CL[pro]. **b** The inhibitory effect of GC376 and boceprevir for SARS-CoV-2 3CL[pro]. **c** The inhibitory effect of identified inhibitors for SARS-CoV-2 3CL[pro].

After protein purification, we evaluated two well-known inhibitors of SARS-COV-2 3CL[pro], namely GC-376 and boceprevir. The calculated $IC_{50}$ values of GC376 and boceprevir for 3CL[pro] were 0.5508 μM and 20.62 μM (Fig. 3b), which were similar to their previously reported $IC_{50}$ values[7]. This result indicates that our enzymatic assay is reliable.

With the established FRET-based assay, the first-round screening was done with the predicted 740 compounds. Encouragingly, there were multiple compounds that showed inhibition against SARS-COV-2 3CL[pro] at 100 μM concentration (Supplementary Fig. 1). Among these compounds, bacitracin, which showed an $IC_{50}$ value of 1.353 μM, was regarded as showing inhibition, and was included in the positive data set, whereas another 739 compounds with $IC_{50}$ values greater than 20 μM were regarded as negative and included in the negative dataset. As described above, we retrained our model and then predicted 17 candidates for the second-round screening. Among them, 10 compounds showed inhibition at 100 μM concentration (Supplementary Fig. 1) and we further tested their binding affinities. The dose-response curves of these compounds were determined under conditions containing

40 μM substrate, 2 μM 3CL$^{pro}$ and different concentrations of tested compounds. There were five compounds exhibiting desirable inhibitory effects in this experiment.

Combining the results of the first and second screening rounds, we have identified six strong inhibitors for SARS-CoV-2 3CL$^{pro}$ (Fig. 3c). Among these molecules, zinc pyrithione inhibited the target enzyme at the lowest concentration in our screening, with an IC$_{50}$ value of 0.9687 μM. It is commonly tested for fungistatic and bacteriostatic use in preclinical research. Similarly, bacitracin has been an FDA-approved drug for years due to its antibiotic activity. Bacitracin also showed strong inhibition of 3CL$^{pro}$ with an IC$_{50}$ of 1.353 μM. More promisingly, Bacitracin can be used as a pediatric medication via intramuscular injection for the systemic treatment of infantile streptococcal pneumonia and empyema. Bacitracin can also be administered as a topical ophthalmic ointment to treat superficial ocular infections involving the conjunctiva and cornea. This shows that both compounds have the potential to work as anti-SARS-CoV-2 medications in clinical treatments.

As a second type of compound, four peptidyl fluorine-methyl-ketones (pFMK), showed potent 3CL$^{pro}$ inhibitory efficiency (IC$_{50}$ below 20 μM), although by varying the peptide sequences, there was a nearly 10-fold difference. This is not surprising, since peptide sequences normally provide the enzyme recognition sites for the molecules. pFMKs are well known covalent caspase inhibitors in biomedical research. While caspases play key roles in apoptosis, pyroptosis and various immune responses, they all share the same active site residue of cysteine with our target enzyme 3CL$^{pro}$. This indicates FMK could be a powerful warhead to react with the active site cysteine of 3CL$^{pro}$ and block its proteolytic activity in a covalent and irreversible manner.

Among the pFMKs, Z-VAD(OMe)-FMK is a well characterized pan-caspase inhibitor that irreversibly binds the catalytic site of various caspases. It prevents cell shrinkage and DNA fragmentation by inhibiting caspase-2, -3, -6, and -8 in flounder immune cells[29] and prevents an increase of p53, PARP-1, and caspase-3 levels in retinal ganglion cells[30]. Through an extension of the peptide chain of Z-VAD(OMe)-FMK with a tyrosine residue, the compound Z-YVAD(OMe)-FMK can block IL-1β secretion, which often initiates hyperinflammation in disease, by inhibiting caspase-1 activity[31]. More recently, Z-WEHD(OMe)-FMK was proven to be a potent, cell-permeable and irreversible caspase-1/5 inhibitor and identified as a robust inhibitor of cathepsin B activity, which is a member of a different cysteine protease family. The known off-target effect of Z-WEHD(OMe)-FMK on cathepsin B might also explain its lower inhibition efficiency for 3CL$^{pro}$ compared to the two former compounds. The last but the most potent pFMK molecule in our assay is Z-IETD(OMe)-FMK, with a IC50 of 1.663 μM. It is a specific caspase-8 inhibitor that disrupts the extrinsic caspase pathway[32], and only partially inhibits the cleavage of caspase-3 and PARP. At non-toxic doses, Z-IETD(OMe)-FMK was found to be immunosuppressive. It was shown to block NF-κB in activated primary T cells but have little inhibitory effect on the secretion of IL-2 and IFN-γ during T cell activation[33]. In the case of SARS-CoV-2, hyperinflammation and the cytokine storm are severe problems after the early stage of the disease. The immunosuppressive function of Z-IETD(OMe)-FMK might help it treat the disease by preventing the damage caused by these more severe effects of the immune response, such as hyperinflammation and the cytokine storm.

**Model interpretation**

As previously mentioned, the mono-fluorinated derivatives are generally irreversible covalent inhibitors of many proteases, which form a covalent thioether adduct with various biological targets including cysteine protease[34–36]. According to this mechanism and some released peptidic covalent reversible/irreversible inhibitors in complex with 3CL$^{pro}$ (e.g., GC376, boceprevir and N3), we assumed that these pFMK inhibitors identified in this study first near to the active site of 3CL$^{pro}$, then their C-terminal warhead (i.e., fluoromethyl ketone) could stable covalently bind to the residue Cys145 of 3CL$^{pro}$ by a nucleophilic attack. To confirm this prediction, we performed a covalent docking following a two-point attractor and flexible side chain methods by AutoDock[37,38]. All of these pFMK inhibitors formed a covalent bond by displacing the fluoride group with the thiolate group of Cys145 (Supplementary Fig. 2). Although covalent inhibitors have become more popular in recent years and shown exciting promise in SARS-CoV-2 therapy[36,39], it is not easy to screen them *in silico*, especially by using docking methods. The process is complicated and time-consuming, and has various limitations. To further complicate matters, large compounds, such as bacitracin, cannot be docked correctly and thus cannot be picked by docking method.

On the contrary, our deep learning model can select most compounds simultaneously regardless of their length and whether they form covalent or non-covalent interactions with the target. However, the results from black-box machine learning model, where humans are not able to understand the process, may confuse people and increase the risk of following false leads, especially in biology and chemistry. Moreover, it is hard to optimize the model and the results without a clear interpretation of the model. Therefore, it is important to map the features extracted by the deep learning model to domain knowledge in order to update our understanding. To explain our model, we explored why our model selects the inhibitors that it does and understand how our model discerns the important sites for binding.

At first, we assumed that our model predicts positive samples due to similar chemical structures in the training set. To test this assumption, we compared the chemical similarity between identified inhibitors and positive compounds in the training set using the Tanimoto similarity score[40]. The Tanimoto nearest neighbor in the training set of Zinc Pyrithione was 1-oxidopyridine-2-thione, a monomer of Zinc Pyrithione. The Tanimoto nearest neighbors of four pFMK inhibitors were Z-FA-FMK and Z-DEVD(OMe)-FMK, which had been reported recently to inhibit 3CL$^{pro}$ with IC$_{50}$ values of 11.39 μM and 6.81 μM, respectively[9]. In that study, Z-FA-FMK inhibited SARS-CoV-2 *in vitro* at the nanomolar level (EC$_{50}$ = 0.13 μM) without apparent cytotoxicity while Z-DEVD(OMe)-FMK did not show potent antiviral activity (EC$_{50}$ > 20 μM). To investigate the influence of nearest neighbors on the predicted results, we removed Z-DEVD(OMe)-FMK and Z-FA-FMK from the training set and re-trained the model. The results show that Z-VAD(OMe)-FMK and Z-IETD(OMe)-FMK were then predicted as negative whereas Z-WEHD(OMe)-FMK and Z-YVAD(OMe)-FMK were still predicted as positive. Presumably, the model does not only rely on the structural similarity, but also relies on hidden features of compounds. To test this hypothesis, we compared the final embedding of compounds from the screening library after dimensionality reduction by t-SNE. As shown in Fig. 4a, Z-FA-FMK, Z-DEVD(OMe)-FMK, and the four identified FMK inhibitors were very close in space after model processing (right, indicated by black arrow) whereas they distributed relatively far before processing (left). These results suggest that the model selects compounds based on their spatial distance to positive compounds in high dimensional space.

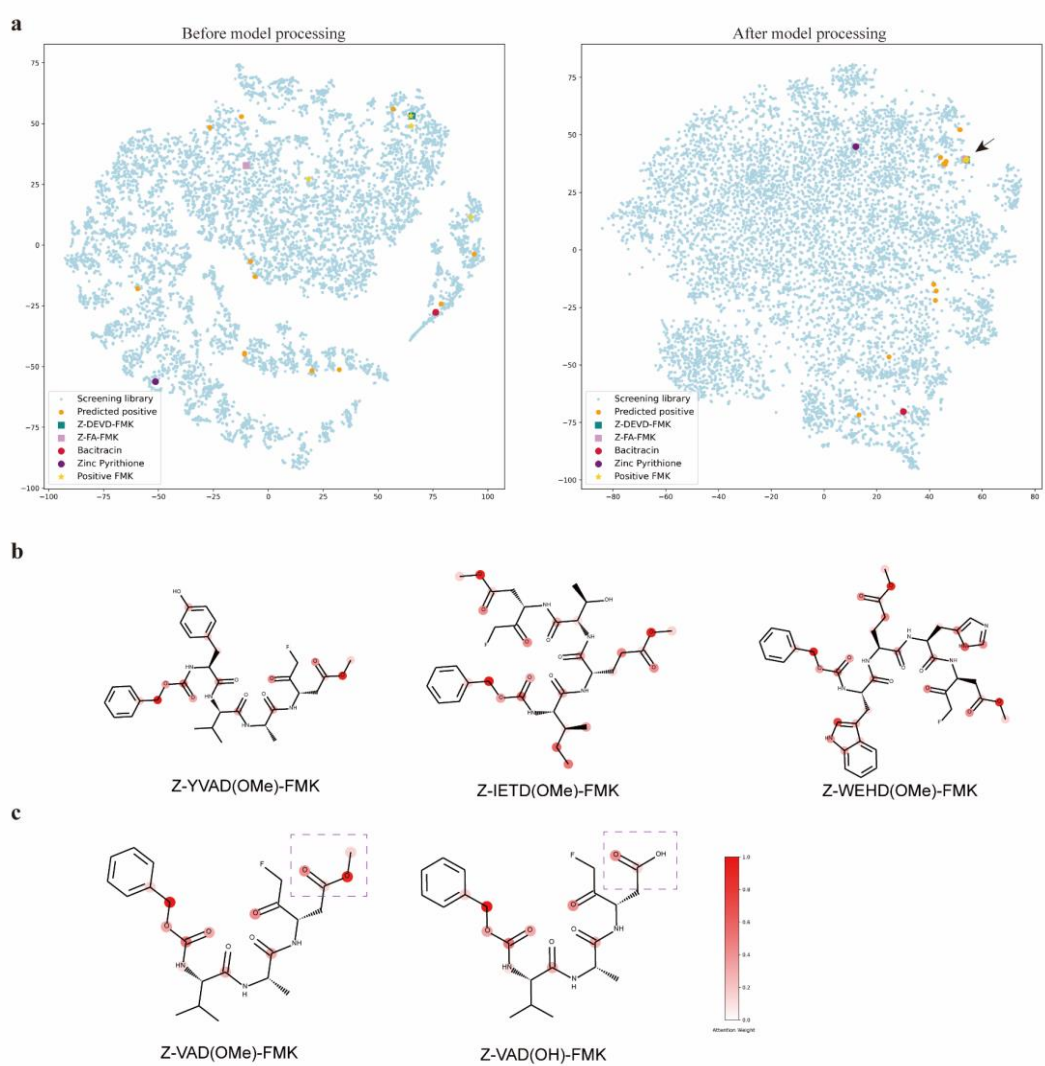

Fig. 4 Visualization and model interpretation. **a** Dimensionality reduction by t-SNE of the molecule representation before model processing (left) and after model processing (right). **b** The attention weights learned by the model indicate important atoms. **c** The difference between Z-VAD(OMe)-FMK (left, predicted positive) and Z-VAD(OH)-FMK (right, predicted negative) is the methylated modification of aspartic acid side chain (indicated by purple dotted lines) and our model thinks this modification is critical for binding.

As mentioned previously, the interactions of other functional groups of these derivatives facilitate the covalent bond between the fluoromethyl ketone warhead and Cys145. It raises the question of whether all pFMK derivatives can bind to 3CL$^{pro}$ or only those with particular features have this binding capability. For our model, the embedding vector of each molecule has its own chemical implications, which has converged information from its atom and bond embeddings through the attention mechanism. To understand this question, we explored why the model considers a compound to be important at the atomic level. We visualized the important atoms ranked by attention weight. As indicated by the color bar (Fig. 4b), the redder in color the atom, the more important it is based on attention weight. We can clearly see some common features across these inhibitors. The carbon near the N-terminal benzene ring (Z) and the carbonyl near the fluoromethyl ketone (FMK) have high attention weights. More importantly, the methylated modification of the last amino acid (D) is considered essential. Interestingly, Z-VAD(OH)-FMK which does not have that methylated modification is predicted as negative. As shown in Fig. 4c, the difference between Z-VAD(OMe)-FMK (predicted positive) and Z-VAD(OH)-FMK (predicted negative) is the methylated modification of the aspartic acid side chain and our model thinks this modification is critical for binding. To prove this hypothesis, we have tested the binding between Z-VAD(OH)-FMK and 3CL$^{pro}$ *in vitro* and calculated the IC$_{50}$ value to be 116.3 μM (Supplementary Fig. 3a). This result suggests that a methylated modification strongly improves the binding ability of these derivatives to 3CL$^{pro}$.

Comparing to IC50 value of 2.554 μM for Z-VAD(OMe)-FMK (Fig.3c), we considered that a methylated modification strongly improves the binding ability of these derivatives to 3CLpro. One possible explanation is that, the methylated modification could significantly reduce molecular polarity and disrupt the molecular interaction force between carboxylic group and cysteine sulfhydryl group of, 3CL$^{pro}$, thus leading to a biologically inactive structure. This is very impressive because our model has successfully captured this important domain knowledge even without any relevant negative samples in the training set. We have also performed docking of Z-VAD(OH)-FMK to 3CL$^{pro}$ and the simulated binding energy is lower than Z-VAD(OMe)-FMK, indicating a stronger binding affinity of Z-VAD(OH)-FMK, as predicted by docking method (Supplementary Fig. 3b). This result suggests that molecular docking cannot capture this critical feature accurately.

Moreover, we have collected and tested all pFMK derivatives from PubChem using our model. For all these 57 derivatives, except six known inhibitors identified in this and previous studies, there are another 10 derivatives that are predicted as positive for 3CL$^{pro}$ (Supplementary Fig. 4). All compounds without the methylated modification of the last amino acid side chain are predicted as negative. These observations indicate that the attention weight of our model has indeed captured domain knowledge at the atomic level. This model interpretation can be used in many applications for mapping to known knowledge and even discovering new knowledge, which may guide drug optimization.

## Biotin-VAD(OMe)-FMK is an activity-based probe of 3CL$^{pro}$

Activity-based probes (ABP) could be used as a chemical antibody to report on the expression of a target protein and have been shown to be remarkable tools due to their ability to label and enrich variable enzymatic activities[41]. An ABP typically consists of three elements: a reactive group (sometimes called a "warhead"), a reporter group (biotin group or fluorophore) and a linker group which could be a peptide sequence used for connecting the two other parts, while functioning as the recognition site for the target enzyme (Supplementary Fig. 5). As mentioned, the covalent bond between the fluoromethyl ketone and Cys145, and the methylated modification of the P1 amino acid (D) were considered important modifications for binding of 3CL$^{pro}$. More interestingly, Biotin-VAD(OMe)-FMK, which was predicted as positive for binding 3CL$^{pro}$, is commercially available. Therefore, we tested whether Biotin-VAD(OMe)-FMK would be a potential ABP for 3CL$^{pro}$. As shown in Supplementary Fig. 6, Biotin-VAD(OMe)-FMK inhibits 3CL$^{pro}$ with an IC$_{50}$ of 6.675 μM, suggesting it is a potent inhibitor of 3CL$^{pro}$. Then we did comparative ABPP experiments using different concentrations of Biotin-VAD(OMe)-FMK to label 3CL$^{pro}$. It was found that the labeling of 3CL$^{pro}$ by Biotin-VAD(OMe)-FMK is remarkably concentration dependent (Fig. 5a). Subsequently, we used the same concentration of Biotin-VAD(OMe)-FMK to incubate with 3CL$^{pro}$ for different times. The results showed that the labeling was also time dependent and a clear band could be visualized even within 5s labeling (Fig. 5b). These experiments verified Biotin-VAD(OMe)-FMK as an activity-based probe (ABP) for 3CL$^{pro}$3CL protease.

Next, we ran competitive ABPP experiments, with Biotin-VAD(OMe)-FMK and four covalent 3CL protease inhibitors we identified, including Z-IETD(OMe)-FMK, Z-YVAD(OMe)-FMK, Z-VAD(OMe)-FMK and Z-WEHD(OMe)-FMK. First, 3CL$^{pro}$ was pre-incubated with different concentrations of inhibitors, then Biotin-VAD(OMe)-FMK was used to label the residual protease[42]. As the concentration of inhibitors increased, the bands gradually became weaker. The results of competitive ABPP are consistent with the fluorescent substrate assay to detect 3CL$^{pro}$ activity. These results suggest that Biotin-VAD(OMe)-FMK could be a powerful activity-based probe for 3CL$^{pro}$ and has a potential application for the labeling of 3CL$^{pro}$ in cell lysate or *in vitro*, even *in vivo*. There are two main advantages of this ABP. First, Biotin-VAD(OMe)-FMK is commercially available. Second, the labeling speed of this probe is very fast. Combined with previous results where Z-VAD(OH)-FMK showed weak inhibition for 3CLpro, these competitive ABPP results prove that the methylated modification of P1 amino acid (D) is truly important for the binding of 3CL$^{pro}$. This finding suggests that model interpretability is absolutely necessary, especially in the application of biology and chemistry, which could help discover new research paths and help us to gain new insights.

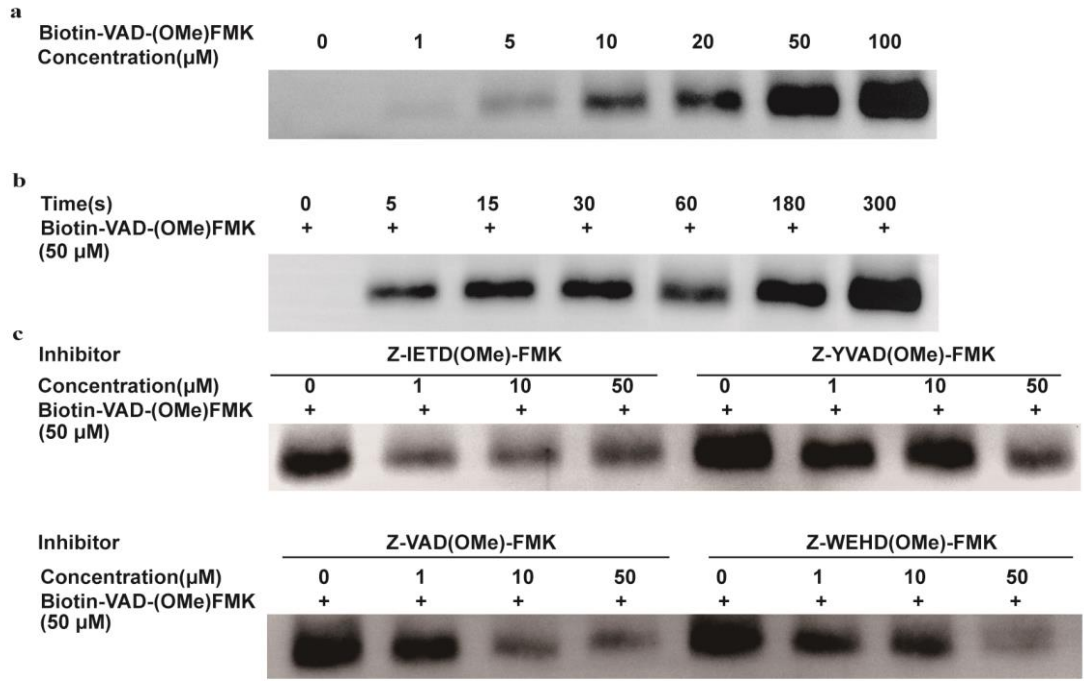

Fig. 5 Comparative and competitive activity-based protein profiling of 3CL$^{pro}$ by Biotin-VAD(OMe)-FMK with Western blot assay. **a** 100 ng 3CL$^{pro}$ was exposed to increasing concentrations of Biotin-VAD(OMe)-FMK for 3 minutes. **b** 100 ng 3CL$^{pro}$ was exposed to 50 μM of Biotin-VAD(OMe)-FMK for different times. **c** 100 ng 3CL$^{pro}$ was pre-incubated with increasing concentrations of four inhibitors for 15 minutes followed by incubation of 50 μM Biotin-VAD(OMe)-FMK for 3 minutes.

# Conclusion

In this work, we propose a framework, namely AIMEE, combining a deep learning model and an enzymatic assay to identify inhibitors against SARS-CoV-2 3CL$^{pro}$. Through this closed-loop and experimental design, our method has achieved excellent performance on a 3CL$^{pro}$ specific highly imbalanced dataset, and has identified and verified six novel potent SARS-CoV-2 3CL$^{pro}$ inhibitors. Among them, there are four outstanding compounds with an IC$_{50}$ value less than 3 μM. More importantly, some of these inhibitors have potential clinical value as COVID-19 therapies. It should be noted that the hit rate of primary screening (100 μM inhibitor) and strong binding (IC$_{50}$ less than 20 μM) in the 2$^{nd}$ round is 58.82% and 29.41%, respectively. This is meaningful not only because of low cost but also because high-throughput screening is not readily available due to BSL3 level of the virus. It proves that building a closed-loop deep learning model combined with wet lab experiments is one of the most effective and promising ways to identify novel inhibitors for a specific target. In addition, it is also possible to easily transfer this 3CL$^{pro}$ model to a homologue target or a protein target which resembles 3CL$^{pro}$ in high dimensional space.

Furthermore, we explored the logic behind the model and evaluated how the model discerns the key sites of the identified compounds. Based on the model interpretation, we have proven that the methylated modification of the FMK nearest amino acid side chain is particularly critical for the binding between the pFMK derivatives and 3CL$^{pro}$. As demonstrated in a previous study, this modification facilitates the binding by avoiding compound cyclization. The mapping of the features captured by a deep learning model to domain knowledge is particularly enlightening, especially for a field relying heavily on wet lab experiments. Since knowledge in such fields has been accumulated over a long period of time and the low-hanging fruit has become so rare, it is now more difficult to get new scientific breakthroughs. Artificial intelligence, particularly deep learning models, may elucidate new paths of research and help us to gain new insights. In this study, we selected a commercially available compound, Biotin-VAD(OMe)-FMK, and proved it could act as an activity-based probe for SARS-CoV-2 3CL$^{pro}$. This opens an avenue for new applications for existing drugs. Taken together, this work highlights the utility of a novel approach incorporating deep learning models and wet lab experiments to expand the boundaries of biological and chemical studies and drug discovery.

## Materials and methods

### Data

The PDBbind dataset, which provides protein-drug complex structures with experimentally measured binding affinities, is commonly used as a benchmark for validating scoring functions [43]. Here PDBbind v2016 core set with diverse 290 complexes covering all protein classes in the PDBbind refined set was used as a test set, for comparisons with other methods. Additionally, CASF-2013 with 195 samples and Astex Diverse Set with 73 samples are also used as independent test sets.

We have collected SARS-CoV/SARS-CoV-2 3CL$^{pro}$ inhibitors (including positive and negative) from various papers [6–9,44] and public datasets such as PubChem, GHDDI and Diamond, as well as validated results in our wet lab experiment. This refined dataset contains carefully selected 3CL$^{pro}$ inhibitors with binding affinity and negative compounds verified in our experiment. We choose an IC$_{50}$ of 20 μM as the threshold to determine positive and negative samples. The refined set then consisted of 408 positive and 1,828 negative samples. The negative set, which contained almost 300 thousand negative compounds for SARS-CoV/SARS-CoV-2 3CL$^{pro}$, was used as a negative sample sampling source for training. After removing overlaps with training set, a collection of 10,924 compounds consisting of approved and clinical-stage drugs, and bioactive compounds was used for *in silico* screening.

All chemicals were of analytical grade and used without further purification. All stock solutions of compounds were prepared in dimethyl sulfoxide (DMSO) and the compounds were purchased from MedChemExpress.

### Model

To fully utilize available information from biological evolutionary relationships and structural information, we explored multi-modal inputs of proteins from different dimensions. Basically, our model consisted of four parts: (i) Protein sequence which were pretrained by large-scale of sequences using masked language model[45], (ii) protein and structure (contact map) data processed by Transformer[26] and Graph Attention Network (GAT)[46], respectively, and then concatenated to get protein embedding vector, (iii) drug smiles that was processed by GAT (Attentive FP model[47]) to get drug embedding vector, (iv) protein and drug embedding vectors were concatenated and fed into fully connected layers to predict interaction.

### Transformer

The most critical part within transformer is multi-head attention[26]. Multi-head attention stacks several modules, namely "scaled dot-product attention", and allows the model to perform parallel computing. The input of each scaled dot-product attention layer consists of queries (q), keys (k) and values (v). Practically, the q, k, v are packed into matrices Q, K and V. Then, the matrix of outputs is calculated as:

$$\text{Attention}(Q, K, V) = \text{softmax}(\frac{QK^T}{\sqrt{d_i}})V \quad (1)$$

Where $d_i$ is the dimension of q and k. Multi-head attention stacks $h$ attention layers and the output matrix is:

$$\text{MultiHead}(Q, K, V) = Concat(head_1, \ldots, head_h)W^O \quad (2)$$

$$\text{where } head_i = \text{Attention}(QW_i^Q, KW_i^K, VW_i^V) \quad (3)$$

Where the projections are parameter matrices $W_i^Q \in R^{d_1 \times d_2}, W_i^K \in R^{d_1 \times d_2}, W_i^V \in R^{d_1 \times d_2}$, and $W^O \in R^{d_1 \times d_1}$, in which $d_1 = hd_2$, and $h$ is the number of heads. We tried different amounts of protein sequence for Transformer pretraining, and finally selected TAPE[48], which was pretrained on 31 million protein domains from Pfam dataset[49]. During training the first 8 layers of this module were frozen.

**Graph Attention Network**

The key point of Graph Attention Network (GAT) is leveraging a self-attention strategy, which was briefly described above, to compute a hidden context of each node by converging its neighbors in the graph[46]. More specifically, for a single graph attentional layer, the representation vector of node i is calculated as:

$$e_{ij} = a(W\vec{h}_i, W\vec{h}_j) \quad (4)$$

$$\alpha_{ij} = softmax(e_{ij}) = \frac{\exp(e_{ij})}{\sum_{k \in N_i} \exp(e_{ik})} \quad (5)$$

$$\vec{h}_i' = \sigma(\sum_{j \in N_i} \alpha_{ij} W\vec{h}_j) \quad (6)$$

Where $\vec{h}_i$, $\vec{h}_j$ are the feature vectors of target node i and neighbor node j. $N_i$ represents all neighbors of node i. W is the trainable weights while $\sigma$ is activate function. Here we use Exponential Linear Unit (ELU) activate function[50]. By aggregating information of every neighbors in the graph through attention, we obtained the representation vector of node i $\vec{h}_i'$.

Then, for $l$ iterations, the GAT was processed in a similar manner to a Graph Neural Network, which involved message passing and readout phases:

$$\vec{h}_i'^{(l-1)} = \sum_{j \in N_i} M^{l-1}(\vec{h}_i^{(l-1)}, \vec{h}_j^{(l-1)}) \quad (7)$$

$$\vec{h}_i^l = GRU^{l-1}(\vec{h}_i'^{(l-1)}, \vec{h}_i^{(l-1)}) \quad (8)$$

Where $M^{l-1}$ and GRU represent graph attention mechanism and activate function gated recurrent unit[51], respectively. $\vec{h}_i^l$ is the representation vector of node i after $l$ iterations, which aggregates the information from node i and its neighbors of iteration $l$-1.

**Focal loss**

We used focal loss to address the negative impact on the model caused by an imbalanced dataset. The focal loss was originally designed to address the one-stage object detection scenario in which foreground is much less than background[28]. Typically, the cross-entropy (CE) loss for a binary classification model (where label is 0 or 1) is defined as:

$$\text{CE} = -(y \log(f(x; w)) + (1 - y) \log(1 - f(x; w))) \quad (9)$$

Where $f()$ and $w$ correspond to the classifier and trainable weights. $x$ and $y$ correspond to the input and label, respectively. For simplification, the predicted probability $p_t$ by the classifier is defined as:

$$p_t = yf(x; w) + (1 - y)(1 - f(x; w)) \quad (10)$$

Thus, CE could be simplified as:

$$\text{CE} = -\log(p_t) \quad (11)$$

For imbalanced dataset, to balance the importance of positive/negative samples and to reduce the impact to gradient by the majority of easily classified negative samples, a weighting factor $\alpha \in [0, 1]$ ($\alpha_t$ is defined similar to $p_t$) and a modulating factor $(1 - p_t)^\gamma$ (tunable focusing parameter $\gamma \geq 0$) were added in CE, thus the focal loss (FL) is defined as:

$$\text{FL} = -\alpha_t (1 - p_t)^\gamma \log(p_t) \quad (12)$$

**Evaluation metrics**

Pearson's correlation coefficient R and root mean square error (RMSE), which refer to the linear correlation and the differences between the predicted and real values, respectively, were used as evaluation metrics:

$$RMSE = \sqrt{\frac{1}{N} \sum_{i=1}^{N} (y_i - \hat{y}_i)^2} \quad (13)$$

$$R = \frac{\sum_{i=1}^{N}(y_i - \bar{y})(\hat{y}_i - \bar{\hat{y}})}{\sqrt{\sum_{i=1}^{N}(y_i - \bar{y})^2} \sqrt{\frac{1}{N}\sum_{i=1}^{N}(\hat{y}_i - \bar{\hat{y}})^2}} \quad (14)$$

where N is the size of a dataset, $y_i$ is the real value whereas $\hat{y}_i$ is the predicted value. $\bar{\hat{y}}$ is the average of real values whereas $\bar{y}$ is the average of the predicted value.

We used three metrics including precision, recall and Matthews correlation coefficient (Mcc) to evaluate the model performance on the imbalanced dataset:

$$Precision = \frac{TP}{TP+FP} \quad (15)$$

$$Recall = \frac{TP}{TP+FN} \quad (16)$$

$$Mcc = \frac{TP \times TN - FP \times FN}{\sqrt{(TP+FP)(TP+FN)(TN+FP)(TN+FN)}} \quad (17)$$

where TP is the number of true positives, TN is the number of true negatives, FP is the number of false positives, FN is the number of false negatives, P indicates positive, and N indicates negative.

**Purification of 3CL$^{pro}$**

A previous study demonstrated that the first serine residue on the N-terminus of SARS-COV 3CL$^{pro}$ is important for its activity and substrate binding and additional residues at the N- and C-terminus of 3CL$^{pro}$ would significantly decrease the enzyme activity[52]. Thus it is important to produce a native 3CL$^{pro}$ for *in vitro* inhibitor screening with no additional residues on either end. Here we chose a novel strategy to obtain a native 3CL$^{pro}$ of SARS-CoV-2. The full gene of SARS-CoV-2 3CL$^{pro}$ was cloned and inserted into a pET28b-SUMO vector, generating a His$_6$-SUMO-tagged fusion protein. Recombinant 3CL$^{pro}$ was expressed in *Escherichia coli* (DE3) and subsequently purified. After digesting with SUMO protease (ULP1), we obtained the native SARS-CoV-2 3CL$^{pro}$ without any redundant residues at either N- or C-terminus (Fig. 3a). In addition, the yield of SARS-CoV-2 3CL$^{pro}$ has been significantly improved (115 mg/L).

**Construction of recombinant plasmid**

The full-length gene encoding SARS-CoV-2 3CL$^{pro}$ (ORF1ab polyprotein residues 3264-3569, GenBank accession number MN908947.3) was optimized and synthesized for Escherichia coli (E. coli) expression (Genscript). All primers and fluorescence substrates were synthesized by Genscript. The synthesized gene was amplified using PCR with the forward primer 5′-CACAGAGAACAGATTGGT GGAAGCGGTTTCCGTAAGATGGCG-3′ and the reverse primer 5′-CTCAGCTTCCTT TCGGGCTTTGTTATTGAAAGGTCACACCGCTGC-3′. This PCR product was employed as the target gene. Then we amplified the vector pET-28b-SUMO with the forward primer 5′-CGCCATCTTACGGAAACCGCTTCCACCAATCTGTTCTCTGTGAG-3′ and the reverse primer 5′-GTGCAGCGGTGTGACCTTTCAATAACAAAGCCCGAAAGGAAGCTG -3′. Through homologous recombination, we connected the target gene with the vector skeleton (ClonExpress II One Step Cloning Kit). At the N-terminus, the construct designed for SARS-CoV-2 3CL$^{pro}$ contains a six-histidine tag (His6-tag) and a SUMO gene (Small Ubiquitin-like Modifier protein).   The C-terminus does not contain any additional amino acids. The native N-terminus was generated after the treatment with SUMO protease, also known as ulp1, which is highly specific for the SUMO protein fusion, recognizing the tertiary structure of SUMO rather than an amino acid sequence.   After the digest with Ulp1, we gained a natural SARS-COV-2 3CL$^{pro}$. The gene sequence of the 3CL$^{pro}$ was verified by sequencing at Sangon Biotech (Shanghai).

**Protein expression and purification**

The sequence-verified SARS-CoV-2 3CL$^{pro}$ construct was transformed into *E. coli* strain BL21-DE3 (Transgene). Transformed clones were pre-cultured at 37°C in 100 mL Luria broth medium with kanamycin (100 μg/mL) overnight, and the incubated culture was inoculated into 1 L Luria broth medium supplied with kanamycin (100 μg/mL) at 37°C. When the cells were grown to OD600 of 0.6-0.8, 0.2 mM isopropyl-D-thiogalactoside (IPTG) was added to the culture to induce the expression of the 3CL$^{pro}$ gene at 16°C. After 20 h, cells were harvested by centrifugation at 4500 x g, 4°C for 10 min. The cell pellets were washed with phosphate-buffered saline twice and then were resuspended in 40 mL lysis buffer (20 mM Tris, 300 mM NaCl, pH 8.0; pH of all buffers was adjusted at room temperature) and then lysed by high-pressure homogenization. The lysate was clarified by ultracentrifugation at 30000 x g at 4°C for 30 min. The supernatant was loaded onto a 5-ml HisTrap HP column (GE Healthcare) equilibrated with lysis buffer containing 20mM imidazole. The HisTrap FF column was washed with 150 mL lysis buffer containing 20mM imidazole to remove nonspecific binding proteins, and eluted with elution buffer (20 mM Tris, 150 mM NaCl, 500 mM imidazole, pH 7.8) with a linear gradient of imidazole ranging from 0 mM to 500 mM, 20 column volumes using an AKTA Pure fast protein liquid chromatography (FPLC). The fractions containing target protein were pooled, then using Amicon Ultra 15 centrifugal filters (10 kD, Merck Millipore) at 4500 x g, 4°C to concentrate the protein. The protein was then mixed with SUMO protease at a molar ratio of 1:100 and dialyzed into reaction buffer (20mM Tris, 100mM NaCl, 1mM DTT, 1mM EDTA; pH 7.8) at 30°C, 1100rpm, for 1 hour. The digested products were loaded onto a Superdex 16/600 size exclusion column equilibrated with elution buffer (330mM Na$_2$HPO$_4$, 167mM NaH$_2$PO$_4$, 150mM NaCl; pH 7.3) and then the eluted fractions were pooled and concentrated.

**SARS-CoV-2 3CL$^{pro}$ kinetic assay**

The standard curve was generated as described below: 2 μM SARS-CoV-2 3CL$^{pro}$ was incubated with varying concentrations of FRET substrate (0.5-40 μM) and the reaction progress was monitored until the fluorescence signals reached plateau, at which point we deemed all the FRET substrate was digested by 3CL$^{pro}$.

For the measurements of Km/Vmax, screening of the protease inhibitor library, as well as IC$_{50}$ measurements, proteolytic reaction with 2 μM SARS-CoV-2 in 50 μL of reaction buffer was carried out at 30 °C in a fluorescence microplate reader with filters for excitation at 360 nm and emission at 460 nm. Reactions were monitored every 40 s. For Km/Vmax measurements, a FRET substrate concentration ranging from 0 to 200 μM was applied. The initial velocity of the proteolytic activity was calculated by linear regression for the first 8 min of the kinetic progress curves. The initial velocity was plotted against the FRET concentration with the classic Michaelis–Menten equation.

## FRET protease assays

The fluorescent substrate harbours the cleave site of SARS-CoV-2 3CL$^{pro}$ and uses Edans and Dabcyl respectively as a donor and quencher pair (Dabcyl-KTSAVLQ↓SGFRKM-E(Edans)-NH$_2$; GenScript). The peptide substrate contains a 14 amino sequence with Dabcyl and Edans attached to its N- and C-terminals, respectively. The fluorescent substrate in a buffer composed of 20mM Tris, 100mM NaCl, 1mM DTT; pH 7.3 was used for the fluorescence resonance energy transfer (FRET) protease assay. Fluorescence readings were obtained using an excitation wavelength of 360 nm and an emission wavelength of 460 nm in a fluorescence microplate reader (BioTek Synergy4) 30 min after the addition of substrate. Initially, we mixed the SARS-CoV-2 3CL$^{pro}$ at the final concentration of 2.0 μM with each compound or DMSO as a negative control in assay buffer and incubated the reaction mixture at 30 °C for 5 min. Next, we added the substrate dissolved in the reaction buffer for a final reaction volume of 50 μL. The final substrate concentrations varied from 5 to 640 μM (5, 10, 20, 40, 80, 160, 320, 640 μM). A calibration curve was generated by measuring different concentrations (from 0.15 to 20 μM) of free Edans in a final volume of 50 μL reaction buffer. Initial velocities were determined from the linear section of the curve, and the relative fluorescence units (RFUs) were determined by subtracting background values (substrate-containing well without protease) from the raw fluorescence values. For the determination of the IC$_{50}$, we incubated 2 μM of SARS-CoV-2 3CL$^{pro}$ with compounds at different final concentrations in reaction buffer at 30 °C for 5 min. Afterwards, we added the FRET substrate to the reaction mixture, for a final concentration of 40 μM and a final total volume of 50 μL to initiate the reaction. Measurements of inhibitory activities of the compounds were performed in triplicate and are presented as the mean ± SD.

## Activity-based protein profiling (ABPP) assay

### Comparative ABPP Assay

For the comparative activity-based protein profiling assay, we prepared 20 μg of total protein for each sample, and made the total volume up to 9 μL by the addition of assay buffer. Next, 1 μL of 10× working stock of Biotin-VAD(OMe)-FMK was added, generating final ABP concentrations of 10 μM, 25 μM, and 50 μM. We then incubated the mixture at 37 °C for different lengths of time (30 s, 1 min and 3 min). 2.5 μL 5× loading buffer was added, and the sample was boiled for 5 min at 100 °C. In this assay, we used deactivated lysate by boiling it with 1% (*wt/vol*) SDS as negative control. The proteins were then separated based on size using a 12% (wt/vol) SDS-PAGE.

### Concentration-dependent experiments

We prepared 100 ng of purified 3CL protease for each sample. The protein solution was premixed with assay buffer, then aliquoted the premixed solution into different tubes. Next, we added 1 μL of 10× working stock of Biotin-VAD(OMe)-FMK, generating final ABP concentrations of 0, 1, 5, 10, 20, 50 and 100 μM. The mixture was incubated at 37 °C for 15 min. Afterwards, the tubes were placed on ice to stop the reaction. Next, we placed the tubes on ice, added 2.5 μL 5×loading buffer to each sample, and boiled the samples for 5 min at 100 °C. The samples were then separated based on size using a 12% (wt/vol) SDS-PAGE.

**Time-dependent experiments**

We prepared 100 ng of purified 3CL protease for each sample. The protein solution was premixed with assay buffer, then aliquoted into different tubes. Next, we added 1 μL of 10× working stock of Biotin-VAD(OMe)-FMK, generating final ABP concentrations of 50 μM. The reaction mixture was incubated at 37 °C for 0 s, 5 s, 15 s, 30 s, 60 s, 180 s and 300 s. Afterwards, we placed the tubes on ice to stop the reaction, added 2.5 μL 5×loading buffer to each sample, and boiled the samples for 5 min at 100 °C. The samples were then run separated based on size using a 12% (wt/vol) SDS-PAGE.

**Competitive ABPP Assay**

We prepared 100 ng of purified 3CL protease for each sample by premixing the concentrated protein solution with assay buffer and then aliquoting it into different tubes. Next, we added different 3CL$^{pro}$ inhibitors Z-IETD(OMe)-FMK, Z-VAD (OMe)-FMK, Z-YVAD(OMe)-FMK and Z-WEHD(OMe)-FMK in the following concentrations: 0, 1,10, and 50 μM. We then incubated the mixture at 37 °C for 15min. Afterwards, we placed the tubes on ice to stop the reaction. Lastly, in order to label the residual active site after inhibition with the 3CL$^{pro}$ inhibitors, we added 1μL 500 μM Biotin-VAD(OMe)-FMK, the activity-based probe, to a final ABP concentration of 50 μM. The samples were then incubated at 37 °C for 3 min.

# Acknowledgements

This work was supported by the National Key Research and Development Program of China (2018YFA0902703), the National Natural Science Foundation of China (NO. 11801542, 31800694 and 31971354), the Strategic Priority Research Program of Chinese Academy of Sciences (NO. XDB 38040200) and the Shenzhen Science and Technology Innovation Committee (JCYJ20170818164014753, JCYJ20170818163445670, JCYJ20180703145002040 and SY8A2211001). We would also like to thank Diana Czuchry for her helpful feedback on a draft of the paper.

# Reference

1. WHO. World Health Organization: Coronavirus disease (COVID-2019) situation reports. 2021; https://www.who.int/emergencies/diseases/novel-cor
2. Zhu N, Zhang D, Wang W, et al. A Novel Coronavirus from Patients with Pneumonia in China, 2019. N. Engl. J. Med. 2020; NEJMoa2001017
3. Wu A, Peng Y, Huang B, et al. Genome Composition and Divergence of the Novel Coronavirus (2019-nCoV) Originating in China. Cell Host Microbe 2020; 27:325–328
4. Lu R, Zhao X, Li J, et al. Genomic characterisation and epidemiology of 2019 novel coronavirus: implications for virus origins and receptor binding. Lancet 2020; 395:565–574
5. Zhang L, Lin D, Sun X, et al. Crystal structure of SARS-CoV-2 main protease provides a basis for design of improved α-ketoamide inhibitors. Science (80-. ). 2020; 368:409–412
6. Jin Z, Du X, Xu Y, et al. Structure of Mpro from SARS-CoV-2 and discovery of its inhibitors. Nature 2020; 582:289–293
7. Fu L, Ye F, Feng Y, et al. Both Boceprevir and GC376 efficaciously inhibit SARS-CoV-2 by targeting its main protease. Nat. Commun. 2020; 11:1–8
8. Ma C, Sacco MD, Hurst B, et al. Boceprevir, GC-376, and calpain inhibitors II, XII inhibit SARS-CoV-2 viral replication by targeting the viral main protease. Cell Res. 2020; 30:678–692
9. Zhu W, Xu M, Chen CZ, et al. Identification of SARS-CoV-2 3CL Protease Inhibitors by a Quantitative High-Throughput Screening. ACS Pharmacol. Transl. Sci. 2020;
10. Krizhevsky A, Sutskever I, Hinton GE. ImageNet classification with deep convolutional neural networks. Commun. ACM 2017; 60:84–90
11. Voulodimos A, Doulamis N, Doulamis A, et al. Deep Learning for Computer Vision: A Brief Review. Comput. Intell. Neurosci. 2018; 1–13
12. Young T, Hazarika D, Poria S, et al. Recent trends in deep learning based natural language processing. IEEE Comput. Intell. Mag. 2018; 13:55–75
13. Schneider P, Walters WP, Plowright AT, et al. Rethinking drug design in the artificial intelligence era. Nat. Rev. Drug Discov. 2020; 19:353–364
14. Wallach I, Dzamba M, Heifets A. AtomNet: A Deep Convolutional Neural Network for Bioactivity Prediction in Structure-based Drug Discovery. Data Min. Knowl. Discov. 2015; 22:31–72
15. Ragoza M, Hochuli J, Idrobo E, et al. Protein-Ligand Scoring with Convolutional Neural Networks. J. Chem. Inf. Model. 2017; 57:942–957
16. Stepniewska-Dziubinska MM, Zielenkiewicz P, Siedlecki P. Development and evaluation of a deep learning model for protein–ligand binding affinity prediction. Bioinformatics 2018; 34:3666–3674
17. Öztürk H, Özgür A, Ozkirimli E. DeepDTA: deep drug–target binding affinity prediction. Bioinformatics 2018; 34:i821–i829
18. Chen L, Tan X, Wang D, et al. TransformerCPI: Improving compound–protein interaction prediction by sequence-based deep learning with self-attention mechanism and label reversal experiments. Bioinformatics 2020;
19. Vellingiri B, Jayaramayya K, Iyer M, et al. COVID-19: A promising cure for the global panic. Sci. Total Environ. 2020; 725:138277
20. Beck BR, Shin B, Choi Y, et al. Predicting commercially available antiviral drugs that may act on the novel coronavirus (SARS-CoV-2) through a drug-target interaction deep learning model. Comput. Struct. Biotechnol. J. 2020; 18:784–790
21. Ton AT, Gentile F, Hsing M, et al. Rapid Identification of Potential Inhibitors of SARS-CoV-2 Main Protease by Deep Docking of 1.3 Billion Compounds. Mol. Inform. 2020; 2000028:1–8
22. Zhang B, Hu Y, Chen L, et al. Mining of epitopes on spike protein of SARS-CoV-2 from COVID-19 patients. Cell Res. 2020; 2–4
23. Stokes JM, Yang K, Swanson K, et al. A Deep Learning Approach to Antibiotic Discovery. Cell 2020; 180:688-702.e13
24. Sanman LE, Bogyo M. Activity-Based Profiling of Proteases. Annu. Rev. Biochem. 2014; 83:249–273
25. Whidbey C, Wright AT. Activity-Based Protein Profiling—Enabling Multimodal Functional Studies of Microbial Communities. Curr. Top. Microbiol. Immunol. 2018; 420:1–21
26. Vaswani A, Shazeer N, Parmar N, et al. Attention Is All You Need. Arxiv 2017;
27. Korkmaz S. Deep Learning-Based Imbalanced Data Classification for Drug Discovery. J. Chem. Inf. Model. 2020;

28. Lin T-Y, Goyal P, Girshick R, et al. Focal Loss for Dense Object Detection. 2017 IEEE Int. Conf. Comput. Vis. 2017; 2999–3007
29. Li S, Li J, Peng W, et al. Characterization of the responses of the caspase 2, 3, 6 and 8 genes to immune challenges and extracellular ATP stimulation in the Japanese flounder (Paralichthys olivaceus). BMC Vet. Res. 2019; 15:20
30. Li G-Y, Fan B, Su G-F. Acute energy reduction induces caspase-dependent apoptosis and activates p53 in retinal ganglion cells (RGC-5). Exp. Eye Res. 2009; 89:581–589
31. Zheng R, Tao L, Jian H, et al. NLRP3 inflammasome activation and lung fibrosis caused by airborne fine particulate matter. Ecotoxicol. Environ. Saf. 2018; 163:612–619
32. Yang J, Pemberton A, Morrison WI, et al. Granzyme B Is an Essential Mediator in CD8 + T Cell Killing of Theileria parva -Infected Cells. Infect. Immun. 2018; 87:
33. Lawrence CP, Chow SC. Suppression of human T cell proliferation by the caspase inhibitors, z-VAD-FMK and z-IETD-FMK is independent of their caspase inhibition properties. Toxicol. Appl. Pharmacol. 2012; 265:103–112
34. Powers JC, Asgian JL, Ekici ÖD, et al. Irreversible Inhibitors of Serine, Cysteine, and Threonine Proteases. Chem. Rev. 2002; 102:4639–4750
35. Citarella A, Micale N. Peptidyl Fluoromethyl Ketones and Their Applications in Medicinal Chemistry. Molecules 2020; 25:4031
36. Cannalire R, Cerchia C, Beccari AR, et al. Targeting SARS-CoV-2 Proteases and Polymerase for COVID-19 Treatment: State of the Art and Future Opportunities. J. Med. Chem. 2020;
37. Morris GM, Huey R, Lindstrom W, et al. AutoDock4 and AutoDockTools4: Automated docking with selective receptor flexibility. J. Comput. Chem. 2009; 30:2785–2791
38. Bianco G, Forli S, Goodsell DS, et al. Covalent docking using autodock: Two-point attractor and flexible side chain methods. Protein Sci. 2016; 25:295–301
39. Ghosh AK, Samanta I, Mondal A, et al. Covalent Inhibition in Drug Discovery. ChemMedChem 2019; 14:889–906
40. Bajusz D, Rácz A, Héberger K. Why is Tanimoto index an appropriate choice for fingerprint-based similarity calculations? J. Cheminform. 2015; 7:20
41. Li N, Overkleeft HS, Florea BI. Activity-based protein profiling: an enabling technology in chemical biology research. Curr. Opin. Chem. Biol. 2012; 16:227–233
42. Li N, Kuo C-L, Paniagua G, et al. Relative quantification of proteasome activity by activity-based protein profiling and LC-MS/MS. Nat. Protoc. 2013; 8:1155–1168
43. Wang R, Fang X, Lu Y, et al. The PDBbind Database: Collection of Binding Affinities for Protein−Ligand Complexes with Known Three-Dimensional Structures. J. Med. Chem. 2004; 47:2977–2980
44. Riva L, Yuan S, Yin X, et al. Discovery of SARS-CoV-2 antiviral drugs through large-scale compound repurposing. Nature 2020;
45. Devlin J, Chang M-W, Lee K, et al. BERT: Pre-training of Deep Bidirectional Transformers for Language Understanding. Arxiv 2018;
46. Veličković P, Cucurull G, Casanova A, et al. Graph Attention Networks. ICLR 2017; 1–12
47. Xiong Z, Wang D, Liu X, et al. Pushing the Boundaries of Molecular Representation for Drug Discovery with the Graph Attention Mechanism. J. Med. Chem. 2019; acs.jmedchem.9b00959
48. Rao R, Bhattacharya N, Thomas N, et al. Evaluating Protein Transfer Learning with TAPE. NIPS 2019; 1–20
49. El-Gebali S, Mistry J, Bateman A, et al. The Pfam protein families database in 2019. Nucleic Acids Res. 2019; 47:D427–D432
50. Clevert D-A, Unterthiner T, Hochreiter S. Fast and Accurate Deep Network Learning by Exponential Linear Units (ELUs). Arxiv 2015;
51. Cho K, van Merrienboer B, Gulcehre C, et al. Learning Phrase Representations using RNN Encoder-Decoder for Statistical Machine Translation. Arxiv 2014;
52. Xue X, Yang H, Shen W, et al. Production of Authentic SARS-CoV Mpro with Enhanced Activity: Application as a Novel Tag-cleavage Endopeptidase for Protein Overproduction. J. Mol. Biol. 2007; 366:965–975